# Positive Correlation between Heavy Alcoholic Drinking and SARS-Cov-2 Not-infection Rate

Ning-Hua Tong[1] and Pietro Salvatori[2]

[1] Department of Physics, Renmin University of China, 100872 Beijing, China

[2] Head & Neck Surgeon, Milan, Italy

[2] Corresponding Author:

Postal address:Via Cesare Saldini, 28 – 20133 Milan, Italy

Email: pietro.salvatori@me.com

mobile phone: +393356645502 Milan, Italy

**COUNT WORDS**: 3014

**ABSTRACT**

**Importance:** During the SARS-CoV-2 pandemic rumours claimed that alcohol drinking could someway be useful in contrasting the contagion and even the disease. This study brings some evidence on this topic.

**Objective:** To determine whether heavy alcohol drinkers and non-drinkers experienced different infection rates.

**Design:** A cohort study done through a simple survey based on the social media software Weixin and the mini survey program Wenjuanxin, was carried out in China, after the zero-Covid policy end, namely from 15:00 Jan.1, 2003, to 12:35 Jan.3, 2023.

**Setting:** The evaluation was conducted among subjects belonging to the first author's Weixin community, most residents in the higher populated China area.

**Participants:** Study participants received a questionary and asked about their virus infection status, and classified it into two groups: (a) infected, meaning he/she has been infected at least once (whether recovered or not); (b) remain uninfected, meaning the virus has not infected him/her. A total of 211 subjects adhered to the survey.

**Exposure:** Alcoholic drinking behaviour about liquors with no less than 40% alcohol content in volume was retrieved from the participants. In China, such beverages are almost uniquely referred to as the Chinese Spirits or BaiJiu. The drinking behaviour was quantified by the frequency of drinking and it is classified into three groups: never drink or drink occasionally (group A); drink one or two times per week (group B); drink three times per week or more often (group C).

**Measures:** The hypothesis of an existing relationship between infection status and drinking behaviour was advanced before data collection. The numbers of the uninfected people in each of the three drinking groups were counted and the rates of not-infection were calculated. The rates are compared with each other to conclude whether significant differences exist, considering the size of the samples. The conclusion is drawn from standard hypothesis testing.

**Results:** Male/female ratio was 108/103 (51.2% and 48.8%), mean age of 38.8 yrs (range 21-68), and median age of 37.4 yrs. We found that 158 (79.4) were infected and 53 (25.1%) were still uninfected at the time of the survey, respectively. The distribution of the three groups with different drinking frequencies was: 139 (65.9%) in group A, 28 (13.3%) in group B, and 44 (20.8) in group C. The statistical analysis of the correlation between uninfected status and drinking behaviour gave these values: versus group B, p=0.63; versus group C, p=0.018; versus group B+C, p=0.048.

**Conclusions and Relevance:** within the limitations of the methodology, this study shows the significative relation between alcohol drinking habits and chances to avoid SARS-CoV-2 infection. The Authors warn about misleading conclusions and advocate research that could properly guide ethanol use in the present and other possible pandemics.

# INTRODUCTION

Coronavirus disease 2019 (COVID-19) has up to now leads to hundreds of millions of infections and more than 6.5 million deaths worldwide [1]. Numerous studies have been devoted to methods that could effectively prevent infection of SARS-Cov-2. Besides the debatable role of vaccination, certain practices in everyday life, such as wearing a mask, social distancing, and washing hands are believed to play positive roles in preventing infection [2]. For example, wearing a mask has been confirmed to be able to reduce the risk of virus transmission [3]. Ethyl Alcohol, or Ethanol (EtOH), has been confirmed to have a strong effect on the virus outside the human body [4, 5]. There are some statements that alcohol intake has no positive effect on preventing the infection of the virus that causes COVID-19 [6]. The negative effect of alcohol drinking on health in the time of the COVID-19 pandemic has been discussed extensively [7]. However, an apparently opposite report found that US counties with high alcohol consumption and high rurality experienced a significantly lower Covid-related mortality rate [8].

In summary, except for the well-known negative effect of alcohol on human health in general, to our knowledge, the correlation between liquor drinking and the rate of infection (or not-infection) of SARS-Cov-2 has not been studied seriously. On the other hand, some reports elucidated the theoretical bases of the EtOH efficacy in eradicating the SARS-Cov-2 from airways [9], the efficacy in preventing infection [10], and improving Covid-19 outcomes [11, 12]. With the end of the dynamic zero-COVID policy in China on Dec.7, 2022, a tide of infection emerges, and the number of people infected with SARS-Cov-2 has increased rapidly in China in January 2023. Most of the infections are proven by antigenic self-test at home. The situation of large number of infections and the availability of self-test kits for the public provides a unique opportunity to study the correlation between the infection rate of SARS-Cov-2 and certain interesting behaviors of everyday life in the population.

In this paper, we report the investigation on the correlation between the infection (or not-infection) rate of SARS-Cov-2 and heavy alcoholic drinking, carried out in a specific time period after the end of the zero-COVID policy and in the restricted population of China. The purpose of the present paper is to report the investigation between the two incidents, i.e., liquor drinking and virus infection, to discuss the eventual correlations and to examine the possible consequences for the public.

**PATIENTS AND METHODS**

A cohort study aimed to investigate the correlation between heavy alcohol drinking and SARS-Cov-2 (referred to as the virus below) infection was carried out in China, by means of a questionary survey. The questionary was generated by the mini program Wenjuanxing, which is included in the social media platform Weixin (the Chinese version of WeChat), and was circulated among the first author's personal contacts, distributed in various Weixin groups, mostly residents in the higher populated China area. Circulation started at 15:00, Jan.1, 2023, and data were collected at 12:35, Jan.3, 2023. The questionnaire was designed to be as simple as possible, concerning only two facts of the investigated individuals, namely: the status of virus infection and the behavior of liquor drinking. All participants were asked to assess their infection status by self-administered antigenic rapid test. As for this variable, participants were classified into two groups: (a) infected, meaning he/she has been infected at least once (whether recovered or not); (b) remain uninfected, meaning he/she has not been infected by the virus. Note that due to the dynamic zero-COVID policy carried out in China before Dec.7, 2022, the number of those who have been infected before that date or have been infected more than once is negligibly small. In this questionnaire, we in practice did not specify and discern the time range of infection and the number of infections.

As for the alcoholic drinking behavior, we only focused on strong liquor which is defined as alcoholic beverages with no less than 40% alcohol content in volume. In China, such beverages are almost uniquely referred to as Chinese Spirits or Baijiu in Chinese pronunciation. The spirit is usually drunk at room temperature, but if warm, this is also acceptable. The drinking behavior is quantified by the frequency of drinking and it is classified into three groups: (group A) never drink or drink occasionally, (group B) drink one or two times per week, and (group C) drink three times per week or more often.

Note that the drinking behavior is quantified by the frequency of drinking instead of the amount of EtOH, since it is believed that, if drinking alcohol is indeed associated with the change of risk in virus infection, it must be the frequency instead of the total amount that is the decisive factor. Also, under the assumption that the content of EtOH could influence the risk of infection - the higher is the content of EtOH, the lower is the risk of infection - we only investigate the strong liquors. Due to the difficulty of accurately estimating the frequency of drinking for many individuals, a more detailed classification of frequency was judged not reliable.

It is hard to estimate the total number of questionnaires distributed since the questionnaire could have been transferred from one Wexin group to others, or from person to person. A crude estimate shows that the percentage of response/distribution is rather low, less than 10%. Finally, a total of 211 questionnaires met the inclusion criteria. All participants gave verbal consent and basic demographics were noted. In order to analyze the correlation between infection status and drinking behavior, the statistical analysis was conducted according to the $\chi^2$ test of goodness of fit and independence.

**RESULTS**

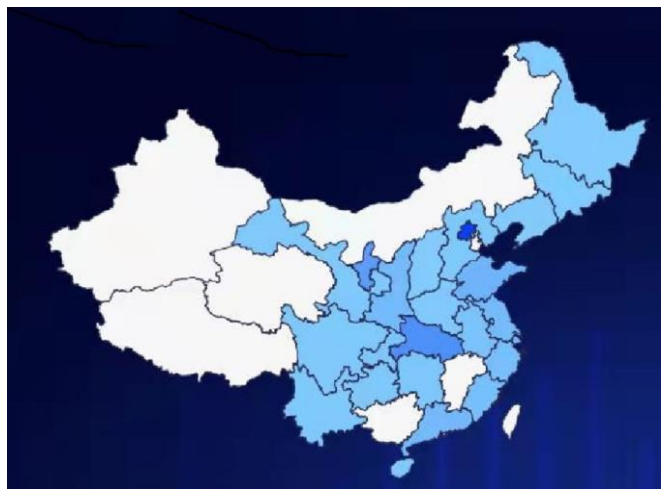

Fig. 1 Geographic distribution of the collected questionnaires in China. Darker blue means a larger number.

Figure 1 shows the geographic distribution of the 211 respondents in China. Although the spread and distribution of the questionnaire are hard to control and are influenced by the social community of the first author, it is seen from Fig.1 that the data are collected from the most heavily populated area of China. The darkest area on the map is Beijing, the city where the first author is working and living. The male/female ratio was 108/103 (51.2% and 48.8%), mean age of 38.8 yrs (range 21-68), and median age of 37.4 yrs.

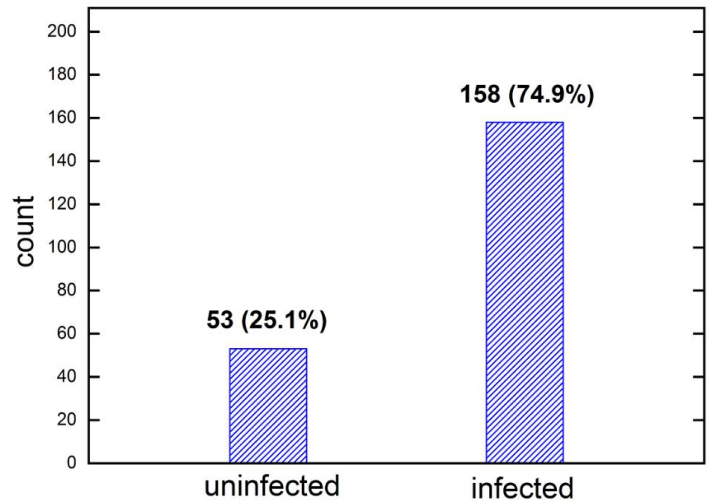

Fig. 2 Count of uninfected people and infected people in the total 211 collected questionnaires. The percentages in the total 211 are shown in the brackets.

Fig.2 shows the distribution according to the infection status: 53 (25.1%) were uninfected and 158 (74.9%) infected.

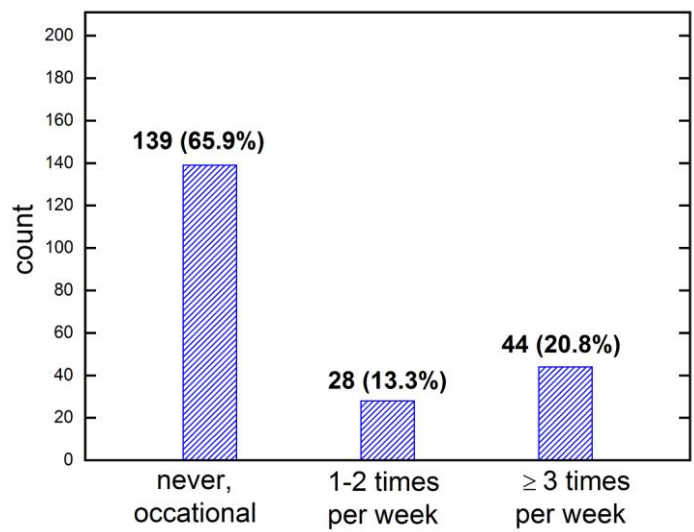

Fig. 3 Count the number of respondents in the three groups with different liquor-drinking frequencies. The percentages in the total 211 are shown in the brackets.

Fig.3 displays the distribution according to drinking behavior: 139 (65.9 %) in group A, 28 (13.3 %) in group B, and 44 (20.8 %) in group C.

**TABLE I:** The count of infected and uninfected respondents in the three drinking groups.

| Drinking Frequency | Never or Occasional | 1-2 times per week | ≥3 times per week | Total |
|---|---|---|---|---|
| Uninfected | 29 (20.9%) | 7 (25.0%) | 17 (38.6%) | **53 (25.1%)** |
| Infected | 110 (79.1%) | 21 (75.0%) | 27 (61.4%) | **158 (74.9%)** |
| **Total** | **139** | **28** | **44** | **211** |

Table 1 summarizes the number of infected and uninfected respondents in each of the three groups with different drinking frequencies. Albeit the number of uninfected respondents varies in each group, the percentage of it in each group shows an interesting trend: it increases from 20.9% in the nondrinker group, to 25.0% in the mild drinker group, and to a much higher value of 38.6% in the heavy drinker group. The average not-infection rate of the total 211 respondents - 25.1% - is close to the value of the mild drinker group.

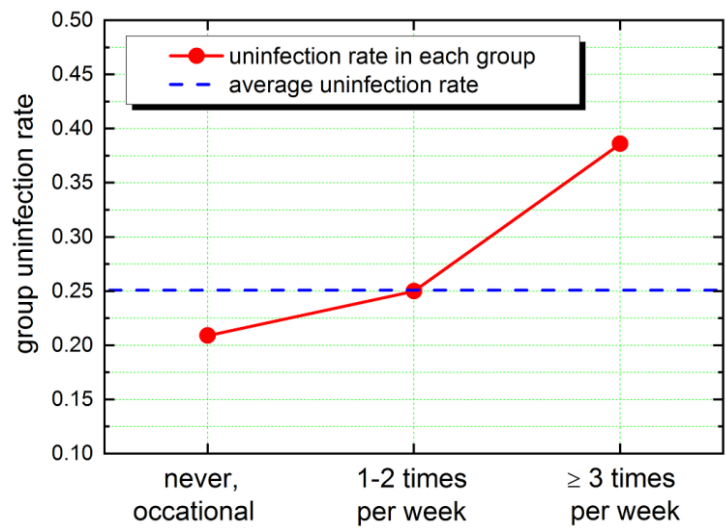

Fig. 4 The non-infection rate as a function of liquor drinking frequency (blue dots). The red dashed line shows the total non-infection rate from the whole sample of 211 respondents.

The trend of the data in Table 1 is shown in Fig.4 as a monotone-increasing curve. This is a surprisingly simple result from the survey: the more frequently people drink liquor, the lower risk they have of being infected with the virus. It is remarkable that the not-infection rate of 38.6% of the heavy drinker group almost doubles the value of 20.9% of the nondrinker group. The statistical analysis of the correlation between uninfected status and drinking behavior gave these values: p=0.63, versus group B; p=0.018, versus group C; p=0.048, versus group B+C.

## DISCUSSION

Overall, Figure 1 shows that the collected data are basically representative of the situation of the whole country.

Before we discuss a possible explanation of the observed significative correlation between heavy liquor drinking and the higher not-infection rate, it appears opportune to dwell on EtOH in the treatment and prevention of Covid-19 in general, in light of recent studies in this field.

First of all, EtOH is a regular drug listed in the USA and EU pharmacopeias and is mainly used for methanol and ethylene glycol poisoning. It has to be remembered that since the 1950s, inhalations of EtOH were proven to be both safe and effective for treating coughs and pulmonary edema. [13, 14]. Moreover, ethanol (up to 9 mg) is frequently used as an excipient in inhalation treatment for asthma and chronic obstructive pulmonary disease [15]. EtOH is also widely used in disinfection procedures. Its antiviral properties derive from the solvent effects on lipids (pericapsid) and from the denaturation of proteins (capsid) [16]. Human coronaviruses, including Severe Acute Respiratory Syndrome Coronavirus (SARS), Middle East Respiratory Syndrome (MERS), Human Endemic Coronavirus, and Influenza-A viruses have been demonstrated to be significantly affected by ethanol on surfaces like plastic and glass, where these viruses can survive for days. Current experimental data show that an ethanol concentration of 30% v/v can inactivate SARS-CoV-2 in 30 seconds [17]. SARS-CoV-2 is an enveloped virus that is extremely sensitive to ethanol, which is also effective against all SARS-CoV-2 variants and other "enveloped" viruses, due to its non-specificity. This particular characteristic broadens the ethanol's range of activity against the SARS-CoV-2 pandemic and suggests its use in potential future epidemics from "enveloped" viruses.

The quantity of EtOH required to reduce the SARS-CoV-2 viral load affecting the lungs was determined by Manning et al. [18] and it amounts to 153 μg or 191.25 μL. Elimination of EtOH occurs at a rate of 120 to 300 mg/L/hour [19]. Alcohol dehydrogenase breaks down 95% of EtOH that has been consumed (or breathed), while the remaining 5% is removed - unaltered - by exhaled air, urine, perspiration, saliva, and tears. Due to the large area of the alveolo-capillar interface, it seems reasonable to assume that 1 fifth of the unaltered, active EtOH escaping the metabolic degradation is eliminated through this pathway. Then, in a normal adult, the amount eliminated through the air is 1% of 120-300 mg/l/hour, so 1.2-3 mg/L over 1 hour. Considering a normal respiratory frequency of 15 acts/min = 900 acts/hour, this means that each exhalation contains 0.0013-0.0033 mg = 1.3-3.3 μg of EtOH. Thus, the calculated dose of 153 μg for inactivating the viral load is approximately exhaled within 118-46 minutes.

Based on the above studies, a possible explanation of our results is that EtOH could reduce the odds of developing infection through a two-way pattern:

1) the EtOH that evaporates while drinking the spirit is inhaled and directly inactivates or destroys the virus lying over the naso-oro-pharyngeal mucosa, which is the most important point of entry for the virus [20], whereas the ingested ethanol that is eliminated through the lungs reaches the upper respiratory tract traveling within the exhaled air. The amount of the exhaled EtOH is probably lower than the inhaled one, but the continuity of respiration grants a longer action over time;

2) the frequent washing of the throat by high concentration alcohol inactivates the virus by direct contact.

It is reasonable to suppose that all 3 events (inhaled, exhaled, direct contact) concur with the final finding in a synergistic way.

Our findings match those showing that US counties with high alcohol consumption and high rurality experienced a significantly lower Covid-related mortality rate [8].

Interestingly, the significance level of the relationship increases with the raising frequency of drinking: absent versus group B (p=0.63), it becomes evident versus group C (p=0.018) and is maintained when considering group B+C (p=0.048). This supports our hypothesis that frequency matters more than the absolute quantity of ingested EtOH. Intuitively, recurrent actuation of the disinfectant improves the chances of inactivating a pathogenic agent.

But there are other explanations as well. For example, it could be that in the group of heavy drinkers, more individuals are in a drunken state all day and their social activity and personal contact are thus significantly reduced, leading to a lower infection rate.

Below, we discuss the prospect of EtOH in the treatment and prevention of Covid-19 in general, in the light of present findings and the literature in this field.

EtOH toxicity is usually – and in some way, improperly – considered an insurmountable obstacle for wider medical usage. First of all, there is a significant difference between ingested and inhaled ethanol, since the latter bypasses the first necessary metabolic step of ingested ethanol and instead travels straight to the left ventricle of the heart and the brain [21]. Second, chronic ethanol use is not the same as chronic ethanol abuse, which can result in lung damage (alveolar macrophage dysfunction, and increased susceptibility to bacterial pneumonia and tuberculosis) [22]. Finally, chronic intoxication has to be differentiated from acute one.

On the topic of acute EtOH inhalation, Bessonneau [23] has demonstrated that the cumulative dose of ethanol inhaled in 90 seconds, while surgically disinfecting hands with a gel containing ethanol at a concentration of 700 g/l, is 328.9 mg.

Rules governing acute ethanol exposure vary by nation or state and are subject to laws. The maximum Blood Alcohol Concentration (BAC) usually ranges between 500 and 800 mg/L.

The regulation also restricts the maximum amount of chronic ethanol exposure in the workplace. For instance, the occupational exposure limit (OEL) for ethanol in the United Kingdom is 1000 parts per million (ppm) of ethanol, or 1910 mg/m3, during an 8-hour shift, which is equivalent to consuming 10 g of ethanol (about one glass of alcohol) daily, according to estimates [24]. These numbers are in perfect agreement with Bessonneau's report [23] and much exceed the amount that would theoretically be needed to reduce the virus load in the respiratory tract [18]. The worries about EtOH inhalation appear to have been completely dispelled by the thorough study of Castro-Balado et al. [25]. Indeed, in rodents breathing, 65% v/v ethanol for 15 min every 8 hours (3 times a day), for five consecutive days (flow rate: 2 L/minute), these Authors examined the possible mucosal or structural damages to EtOH in the lung, trachea, and esophagus. In this experiment, the calculated absorbed dosage was 1.2 g/kg/day. Under the same conditions, this dosage in humans would be equivalent to 151 g per day. Notably, neither the treated animals nor the controls' histology samples showed any signs of damage. A recent RCT from the same group has confirmed these data in humans [12].

Interestingly, in the RCT from Hosseinzadeh [10], no collateral effects have been mentioned, perhaps because were lacking or minimal and tolerable, whereas Amoushahi et al. reported few and bearable adverse events [11]. Finally, numerous studies suggest that industrial exposure is not a problem in reproductive medicine (Irvine) [26], nor in oncology (Bevan) [24], despite the toxicity of chronic ethanol inhalation.

In our opinion, papers intended to reasonably combat the inappropriate EtOH use in Covid-19 seem not to have taken due account the knowledge of EtOH, including toxicity. Indeed, the document released by WHO [6] is merely divulgation and does not bring any evidence supporting the section "General myths about alcohol and COVID-19", but simply affirms that "there is no evidence that drinking alcohol offers any protection against COVID-19 or has a positive effect on the course and outcomes of any infectious disease". Now, if this might have been acceptable in 2020, at present the evidence showed by Pro et al. [8] should lead to a selective update, at least. Particularly, the RCT from the Spanish group [12] found a faster and higher - even not significative - reduction of the viral load in the EtOH-treated patient's group versus controls. Unfortunately, the RCT had to be ended earlier for insufficient recruitment.

It should also be remembered that aerosol delivery is more efficient than simple inhalation. Moreover, in the editorial from "Alcohol and alcoholism" [7] the Author reports a considerable number of deaths from "alcohol" intake during the pandemic. But, when examining the cited references, it appears clear that they refer either to methanol, or other drugs added to "alcoholic beverages". Moreover, these papers report events that occurred in 2018 and 2019, then before the SARS-CoV-2 pandemic.

Due to these someway misleading messages, the medical body and health authorities could have neglected an efficacious, easily available treatment to help pandemic control.

Prof. Shintake [27] on March 17, 2020, and Dr. Amoushahi et al. [28] on May 25, 2020, are credited for first hypothesizing EtOH treatment to prevent or eradicate SARS-Cov-2 infection. Today, it seems there is sufficient research body leading to deeply test the EtOH efficacy and efficiency on airways disinfection in SARS-CoV-2-positive subjects and Covid-19 patients [9, 10, 11, 12]. The positive correlation between heavy alcohol drinking and the SARS-Cov-2 not-infection rate reported in the present paper lends itself to the importance of such research.

This survey investigation could have many flaws and sources of error. They include but are not limited to:

1. pollution of sample: repeated or dishonest submission of questionnaire; inaccuracy in the Antigen self-test result;

2. the bias of sample: the investigation is limited to active users of the social media platform Weixin, effectively expelling younger, elder, seriously sick, and deceased individuals;

3. the relatively small size of the sample;

4. correlations in the sample: the infection of family members and friends could be correlated; the not-infection rate may be correlated to the liquor drinking behavior through variables such as age, class, living area, etc.

For the first factor, the mini program Wenjuanxing provides each respondent's IP address, and one can exclude the repeated submission to some extent. But for most of the factors, it is hard to evaluate their influence on the conclusion of the present investigation.

It has to be made clear that the Authors are deeply aware of the harmful potential of the EtOH and do not absolutely support its oral intake for the prevention or treatment of SARS-CoV-2 infection and Covid-19 disease. However, they believe that a more reliable word on this topic is necessary for the correct public health management, either to offer an efficacious/efficient treatment or to avoid possible damages from the EtOH myth or misuse. Because Science does not fight Myth by replacing it with another Myth.

**ACKNOWLEDGMENTS**

N.-H. Tong thanks Min Luo, Ze-Dong Lin, and Zhi- Yuan Xie for help in the survey investigation and for giving suggestions. He also thanks Wei Liu and Johannes Ni for valuable information and insightful discussions. The Authors have not conflict of interests nor funding sources.

**REFERENCES**

1. World Health Organization. Coronavirus disease (COVID-2019) situation reports. Weekly epidemiological update on COVID-19, 26 October 2022.

2. Moreland S, Zviedrite N, Ahmed F, and Uzicanin A. COVID-19 Prevention at Institutions of Higher Education, United States, 2020 2021: Implementation of Nonpharmaceutical Interventions, medRxiv 2022.07.15.22277675; doi: https://doi.org/10.1101/2022.07.15.22277675.

3. Leech G, Rogers-Smith C, Sandbrink JB, Snodin B, Zinkov R, Rader B, Brownstein JS, Gal Y, Bhatt S, Sharma M, Mindermann S, Brauner JM, and Aitchison L, Mass mask-wearing notably reduces COVID-19 transmission, medRxiv 2021.06.16.21258817; doi: https://doi.org/10.1101/2021.06.16.21258817.

4. Singh D, Joshi K, Samuel A, Patra J, and Mahindroo N. Alcohol-based hand sanitisers as first line of defence against SARS-CoV-2: a review of biology, chemistry and formulations, Epidemiology and Infection 148, e229, 19. https: doi.org/10.1017/S0950268820002319.

5. Nomura T, Nazmul T, Yoshimoto R, Higashiura A, Oda K, and Sakaguchi T. Ethanol susceptibility of SARS-CoV-2 and other enveloped viruses, Biocontrol Science, 2021, Vol. 26, No. 3, 177180, doi: 10.4265/bio.26.177.

6. https://www.euro.who.int/__data/assets/pdf_file/0010/437608/Alcohol-and-COVID-19-what-you-need-to-know.pdf.

7. Chick J, Editorial, Alcohol and COVID-19, Alcohol and Alcoholism, 2020, 55(4), 341342, doi:10.1093/alcalc/agaa039.

8. Pro G, Gilbert PA, Baldwin JA, Brown CC, Young S, Zaller N (2021) Multilevel modeling of county-level excessive alcohol use, rurality, and COVID-19 case fatality rates in the US. PLoS ONE 16(6): e0253466. https://doi.org 10.1371/journal. pone.0253466.

9. Salvatori P. The rationale of ethanol inhalation for disinfection of the respiratory tract in SARS-CoV-2-positive asymptomatic subjects. Pan African Medical Journal. 2021;40:201. [doi: 10.11604/pamj.2021.40.201.31211]

10. Hosseinzadeh A, Tavakolian A, Kia V, Ebrahimi H, Sheibani H, Binesh E, Jafari R, Mirrezaie SM, Jafarisani M, and Emamian MH. Application of nasal spray containing dimethyl sulfoxide (DMSO) and ethanol during the COVID-19 pandemic may protect healthcare workers: A randomized controlled trial. Iran Red Crescent Med J. 2022 August; 24(8):e1640 doi: 10.32592/ircmj.2022.24.8.1640

11. Amoushahi A, Moazam E, Tabatabaei AR, Ghasimi G, Grant-Whyte I, Salvatori P, Ezz AR. Efficacy and Safety of Inhalation of Nebulized Ethanol in COVID-19 Treatment: A Randomized Clinical Trial. Cureus. 2022 Dec 5;14(12):e32218. doi: 10.7759/cureus.32218.

12. Castro-Balado A, Novo-Veleiro I, Vázquez-Agra N, Barbeito-Castiñeiras G, Estany-Gestal A, Trastoy-Pena R, González-Barcia M, Zarra-Ferro I, del Río-Garma MC, Crespo-Diz C, Delgado-Sánchez O, Otero-Espinar FJ, Mondelo-García C, Pose-Reino A, Fernández-Ferreiro A. Efficacy and Safety of Inhaled Ethanol in Early-Stage SARS-CoV-2 Infection in Older Adults: A Phase II Randomized Clinical Trial. Pharmaceutics. 2023; 15(2):667. https://doi.org/10.3390/pharmaceutics15020667

13. Gootnick A, Henry I. Lipson HI, and Joseph Turbin J. "Inhalation of ethyl alcohol for pulmonary edema." New England Journal of Medicine, November 29, 1951, 245: 842-843

14. Burnham PJ.Q. A new rapid treatment for the useless cough in the postoperative patient. Bull Northwest Univ Med Sch. 1954;28(1):76-8.

15. https://www.sps.nhs.uk/artcles/ethanol-content-of-inhalers-what-is-the-signifcance/

16. Kampf G. Efficacy of ethanol against viruses in hand disinfection, Journal of Hospital Infection, 98 (2018) 331-338. http://dx.doi.org/10.1016/j.jhin.2017.08.025.

17. Kratzel A, Todt D, V'kovski P, Steiner S, Gultom M, Nhu Thao TT, et al. Inactivation of Severe Acute Respiratory Syndrome Coronavirus 2 by WHO-Recommended Hand Rub Formulations and Alcohols. Emerg Infect Dis. 2020 Jul; 26(7): 1592–1595. https://pubmed.ncbi.nlm.nih.gov/32284092/

18. Manning TJ, Thomas-Richardson J, Cowan M, Thomas-Richardson G. Should ethanol be considered a treatment for COVID-19? Rev Assoc Med Bras. 2020 Sep;66(9):1169-71. https://pubmed.ncbi.nlm.nih.gov/33027439/

19. Winek CL, Murphy KL. The rate and kinetic order of ethanol elimination. Forensic Sci Int. 1984;25(3):159-66. https://pubmed.ncbi.nlm.nih.gov/6745819/

20. Sungnak W, Huang N, Bécavin C, Berg M, Queen R, Litvinukova M, et al. SARS-CoV-2 entry factors are highly expressed in nasal epithelial cells together with innate immune genes, Nat Med. 2020;26:681–7 https://doi.org/10.1038/s41591-0200868-6

21. MacLean RR, Valentine GW, Jatlow PI, Sofuoglu M. Inhalation of Alcohol Vapor: Measurement and Implications. Alcohol Clin Exp Res. 2017 Feb; 41(2): 238–250. Published online 2017 Jan 5. doi:10.1111/acer.13291

22. Yeligar SM, Chen MM, Kovacs EJ, Sisson JH, Burnham EL, and Brown LAS. Alcohol and Lung Injury and Immunity, Alcohol. 2016 Sep; 55: 51–59. doi:10.1016/j.alcohol.2016.08.005

23. Bessonneau V, Thomas O. Assessment of Exposure to Alcohol Vapor from Alcohol-Based Hand Rubs, Int J Environ Res Public Health. 2012 Mar;9(3):868–79. https://www.ncbi.nlm.nih.gov/pmc/articles/PMC3367283/

24. Bevan RJ, Slack RJ, Holmes P, Levy LS. An assessment of potential cancer risk following occupational exposure to ethanol J Toxicol Environ Health B Crit Rev. 2009 Mar;12(3):188-205. https://pubmed.ncbi.nlm.nih.gov/19466672/.

25. Castro-Balado A, Mondelo-Garcia C, Barbosa-Pereira L, Varela-Rey I, Novo-Veleiro I, Vazquez Agra N, et al. Development and Characterization of Inhaled Ethanol as a Novel Pharmacological Strategy Currently Evaluated in a Phase II Clinical Trial for Early-Stage SARS-CoV-2 Infection. Pharmaceutics. 2021;13:342. https://doi.org/10.3390/ pharmaceutics13030342

26. Irvine LFH. Relevance of the developmental toxicity of ethanol in the occupational setting: a review, J Appl Toxic. 2003 Sep-Oct;23(5): 289-99. https://analyticalsciencejournals.onlinelibrary.wiley.com/doi/10.1002/jat.937

27. Shintake T. Possibility of Disinfection of SARS-CoV-2 (COVID-19) in Human Respiratory Tract by Controlled Ethanol Vapor Inhalation, arXiv:2003.12444v1 [physics.med-ph]

28. Amoushahi A, Padmos A. A Suggestion on Ethanol Therapy in COVID-19? https://www.ecronicon.com/ecan/pdf/ECAN-06-00229.pdf